# Compact Securities Markets for Pareto Optimal Reallocation of Risk

David M. Pennock
NEC Research Institute
4 Independence Way
Princeton, NJ 08540
dpennock@research.nj.nec.com

Michael P. Wellman
University of Michigan AI Laboratory
1101 Beal Avenue
Ann Arbor, MI 48109-2110 USA
wellman@umich.edu

## Abstract

The *securities market* is the fundamental theoretical framework in economics and finance for resource allocation under uncertainty. Securities serve both to reallocate risk and to disseminate probabilistic information. *Complete* securities markets—which contain one security for every possible state of nature—support Pareto optimal allocations of risk. Complete markets suffer from the same exponential dependence on the number of underlying events as do joint probability distributions. We examine whether markets can be structured and "compacted" in the same manner as Bayesian network representations of joint distributions. We show that, if all agents' risk-neutral independencies agree with the independencies encoded in the market structure, then the market is *operationally complete*: risk is still Pareto optimally allocated, yet the number of securities can be exponentially smaller. For collections of agents of a certain type, agreement on Markov independencies is sufficient to admit compact and operationally complete markets.

## 1  INTRODUCTION

A large portion of the world's economic transactions involve the exchange of risk. For example, insurance policy holders transfer some of their risks to insurance providers, in exchange for sure payments. Farmers hedge against the dangers of adverse weather by exchanging futures with less risk-averse speculators. Insurance contracts, futures, options, derivatives, and even stocks, serve to continuously reallocate risk around the globe.

All of these potentially complex financial instruments can be modeled as portfolios of much simpler instruments, called *securities*. Securities are essentially lottery tickets: they pay off in some good (e.g., money) contingent on the outcomes of uncertain events. A key result in the theory of economics under uncertainty is that, if agents have access to "enough" securities (i.e., access to a *complete market*), then equilibrium allocations of risk are Pareto optimal. Unfortunately, "enough" is, for all intents and purposes, too much: the number of required securities is equal to the size of the *joint* space of all relevant uncertain events, and is thus intractable in any realistic setting.

The prospect of representing probabilities over joint event spaces was once viewed in much the same light—theoretically ideal, but practically unachievable. The advent of graphical modeling languages, and in particular Bayesian networks (BNs), changed this view dramatically. These languages permit concise descriptions of otherwise unwieldy joint distributions, as long as sufficient conditional independencies among events are present. In this paper, we demonstrate that, among certain populations of agents, conditional independence can be analogously exploited in the design and configuration of securities markets.

Section 4 shows how securities markets can be structured according the the topology of any BN. As with BNs, if sufficient independencies are encoded in the structure, the size of the market is exponentially reduced. Although structured markets are not complete in the traditional sense, we derive conditions under which they are nonetheless *operationally complete*, meaning that the equilibrium allocation of risk is still Pareto optimal. Section 5.1 gives a general sufficient condition: if, in equilibrium, all agents' risk-neutral independencies agree with those encoded in the market's structure, then the market is operationally complete. Section 5.2 characterizes the computational complexity of pricing securities and finding arbitrage opportunities in a structured market. Section 6 derives a special case when agreement on true independencies is sufficient to yield operationally complete markets; we also explain why agreement on true independencies is not sufficient in general.

## 2 BACKGROUND AND NOTATION

We consider a model economy of $N$ agents, indexed $i = 1, 2, \ldots, N$, each with a subjective probability distribution $\Pr_i$ over states of the world and a utility function $u_i$ for money. Denote the set of all possible states of the world as $\Omega = \{\omega_1, \omega_2, \ldots\}$. The $\omega$ are mutually exclusive and exhaustive.

State is often more concisely and naturally characterized as the set of outcomes of *events*. Denote the set of modeled events as $Z = \{A_1, A_2, \ldots, A_M\}$. Underlying $M$ arbitrary events is a state space $\Omega$ of size $|\Omega| = 2^M$, consisting of all possible combinations of event outcomes. Conversely, any set of states can be factored into a set of $M = \lceil \lg |\Omega| \rceil$ events. Without further assumption, the two representations are equivalent in both expressivity and size, although the event factorization may be more natural. In most of what follows, the events $\{A_j\}$ are the focus of attention, with $\Omega$ the implied joint outcome space. We refer to the $\{A_j\}$ as the *primary events*, so as to distinguish them from the other $2^{2^M} - M$ possible sets of states, each of which is also an event.

### 2.1 DECISION MAKING UNDER UNCERTAINTY

In general, an agent's utility is defined over the cross product of available actions and possible states. We assume here that utility arises from an underlying utility for *money*. If agent $i$'s utility for $\mu$ dollars is $u(\mu)$, then its utility $U$ for a particular action $a$ is its expected utility for money,

$$U_i(a) = E_i\left[u_i\left(\Upsilon_i^{\langle\omega\rangle}\right)\right] = \sum_{\omega \in \Omega} \Pr_i(\omega) u_i\left(\Upsilon_i^{\langle\omega\rangle}\right), \quad (1)$$

where $\Upsilon_i^{\langle\omega\rangle}$ is agent $i$'s wealth in dollars when action $a$ is taken in state $\omega$ (the dependence of $\Upsilon_i^{\langle\omega\rangle}$ on $a$ is implicit). Agent $i$'s *decisions* are made by maximizing expected utility, or choosing the action $a$ that maximizes (1).

We assume throughout that utility increases monotonically with wealth. *Local risk aversion* at $\mu$, denoted $r_i(\mu)$, is defined as $r_i(\mu) \equiv -u_i''(\mu)/u_i'(\mu)$. Agent $i$ is *risk-averse* if $r_i(\mu) > 0$ for all $\mu$, or, equivalently, if $u_i$ is everywhere concave. Under this condition, the agent always prefers a guaranteed payment equal to the expected value of a lottery rather than the lottery itself, thus exhibiting an "aversion" to gambling. The agent is *risk-neutral* if $r_i(\mu) = 0$ for all $\mu$, or $u_i$ is linear; in this case, maximizing (1) coincides with maximizing expected payoff.

### 2.2 RISK-NEUTRAL PROBABILITY

Notice that an outside observer $O$, privy only to agent $i$'s chosen actions, cannot uniquely discern either the agent's belief or its utility: the two quantities are inextricably linked (Kadane & Winkler, 1988). Any one of a continuous family of belief–utility pairs offers an equally valid rationalization for the agent's actions. That is, for any function $f(\omega)$, subjective probabilities proportional to $\Pr_i(\omega)f(\omega)$ matched with utilities $u_i\left(\Upsilon_i^{\langle\omega\rangle}\right)/f(\omega)$ result in strategically equivalent utilities for actions $U_i(a)$.

*Risk-neutral* probabilities are defined as

$$\Pr_i^{\text{RN}}(\omega) \propto \Pr_i(\omega) u_i'\left(\Upsilon_i^{\langle\omega\rangle}\right), \quad (2)$$

where $u_i'$ is the derivative of utility (Nau, 1995). Agent $i$'s observable behavior, manifested as actions, is indistinguishable from that of a hypothetical agent with transformed probabilities $\Pr_i^{\text{RN}}(\omega)$ and reciprocally transformed utility $u_i^{\text{RN}}(\mu) \equiv u_i(\mu)/u_i'\left(\Upsilon_i^{\langle\omega\rangle}\right)$. It turns out that the observer *can* uniquely assess agent $i$'s risk-neutral probabilities. In fact, all standard elicitation procedures designed to reveal agent $i$'s beliefs based on monetary incentives (de Finetti, 1974; Winkler & Murphy, 1968)—for example, querying the prices at which the agent would buy or sell various lottery tickets—essentially reveal $\Pr_i^{\text{RN}}$, and *not* $\Pr_i$ (Kadane & Winkler, 1988). The agent's *observable* beliefs are in effect its risk neutral probabilities, not its true probabilities.

### 2.3 SECURITIES MARKETS FOR THE REALLOCATION OF RISK

Under uncertainty, risk-averse agents will desire to *hedge* or *insure* against their risks by distributing wealth across states. For example, insuring the delivery of a package effectively transfers wealth from the *package-received* state to the *package-lost* state. The Arrow-Debreu securities market is the fundamental theoretical framework in economics and finance for resource allocation under uncertainty (Arrow, 1964; Dreze, 1987; Mas-Colell, Whinston, & Green, 1995). A *security*, denominated in money or other exchangeable good, pays off variously contingent upon the realization of an uncertain state. Let $\langle A \rangle$ denote a security that pays off one dollar if and only if the event $A$ occurs. If the price of this security is $p^{\langle A \rangle}$ per unit, then agent $i$'s decision to purchase $x_i^{\langle A \rangle}$ units is equivalent to accepting a lottery with payoff $(1 - p^{\langle A \rangle})x_i^{\langle A \rangle}$ if $A$ occurs, and $-p^{\langle A \rangle}x_i^{\langle A \rangle}$ otherwise. Positive $x_i^{\langle A \rangle}$ indicates a quantity to buy, and negative $x_i^{\langle A \rangle}$ a quantity to sell.

In a market of $S$ such securities, let $\mathbf{p} = \langle p^{\langle 1 \rangle}, p^{\langle 2 \rangle}, \ldots, p^{\langle S \rangle} \rangle$ denote the securities' prices, and $\mathbf{x}_i = \langle x_i^{\langle 1 \rangle}, x_i^{\langle 2 \rangle}, \ldots, x_i^{\langle S \rangle} \rangle$ denote the quantities of the securities held by agent $i$. Agent $i$'s utility for securities is its expected utility for money (1), where the agent's choice of actions is how much to buy or sell of each security.

Agents trade securities with each other prior to revelation of the world state. In an economy of $N$ agents, each continually maximizing (1), prices adjust until all buy orders match with sell orders for all securities. A market is in *competitive equilibrium* at prices **p** if and only if

$$\sum_{i=1}^{N} \mathbf{x}_i(\mathbf{p}) = \mathbf{0}, \quad (3)$$

where $\mathbf{x}_i(\mathbf{p})$ is agent $i$'s optimal demand vector at prices **p**.

A securities market is termed *complete* if it contains at least $|\Omega|-1$ *linearly independent* securities. Such a market guarantees, under classical assumptions, that equilibrium entails a *Pareto optimal*, or *efficient*, allocation of risk.

A *conditional security* $\langle A_1 | A_2 \rangle$ pays off *contingent* on $A_1$ and *conditional* on $A_2$. That is, if $A_2$ occurs, then it pays out exactly as $\langle A_1 \rangle$; on the other hand, if $\bar{A}_2$ occurs, then the bet is called off and any price paid for the security is refunded (de Finetti, 1974). The canonical complete market consists of one security paying out in each state of nature. In general, though, any set of securities (possibly including conditionals) with a payoff-by-state matrix of rank $|\Omega|-1$ is complete.

When one unit of each security pays out one dollar, the equilibrium prices in a securities market form a coherent probability distribution. For example, $p^{\langle A_1 \rangle} = p^{\langle A_1 A_2 \rangle} + p^{\langle A_1 \bar{A}_2 \rangle}$; or $p^{\langle A_1 A_2 \rangle} = p^{\langle A_1 | A_2 \rangle} p^{\langle A_2 \rangle}$. In fact, the equilibrium prices coincide with the agents' risk-neutral probabilities (2) for the available securities, which must be in complete agreement (Dreze, 1987; Nau & McCardle, 1991). Derived formally in Section 3.1, we simply sketch the intuition here. Since a risk-neutral agent buys $\langle A_j \rangle$ if $p^{\langle A_j \rangle} < \Pr_i(A_j)$ (it simply maximizes expected payoff), then *any* agent buys $\langle A_j \rangle$ if $p^{\langle A_j \rangle} < \Pr_i^{RN}(A_j)$. Similarly, the agent sells if $p^{\langle A_j \rangle} > \Pr_i^{RN}(A_j)$. If two agents $h$ and $i$ have differing risk neutral probabilities—that is, $\Pr_h^{RN}(A_j) \neq \Pr_i^{RN}(A_j)$—then there is an intermediate price at which they are both willing to trade. It follows that, at equilibrium, when by definition opportunities for exchange have been exhausted, all agents' risk neutral probabilities agree across available securities. Furthermore, since offers to buy and sell must match, the equilibrium prices equal these consensus probabilities.

There are two, largely inseparable, reasons for agents to trade in securities: to insure against risk ("hedge") and to profit from perceived mispricings ("speculate"). The more averse to risk, the more the former consideration dominates an agent's decision making. On the other hand, risk-neutrality—the limit of diminishing risk aversion—is synonymous with pure speculation. These two behaviors are aligned with the two central roles of securities markets in the theory of economics under uncertainty. The first, as mentioned, is to support the reallocation of risk. The second is to *aggregate* and *disseminate* information. Agents that disagree on the likelihood of states may seek to exchange securities at prices that yield, according to each's subjective viewpoint, an increase in expected returns. Moreover, each agent is privy, albeit implicitly, to the evidence gathered by other agents (perhaps at great cost) via fluctuations in price.

### 2.4 BAYESIAN NETWORKS

A joint probability distribution can often be represented more compactly as a *Bayesian network (BN)*, or other graphical model (Darroch, Lauritzen, & Speed, 1980). Conciseness is achieved by exploiting conditional independence among the primary events. Let $\text{CI}[A_j, W, X]$ be shorthand for $\Pr(A_j|WX) = \Pr(A_j|W)$, indicating that $A_j$ is conditionally independent of the set of events $X$, given another set $W$. Consider the event $A_k \in Z$, with predecessors $\mathbf{pred}(A_j) \equiv \{A_1, A_2, \ldots, A_{k-1}\}$. Suppose that, given the outcomes of a subset $\mathbf{pa}(A_k) \subseteq \mathbf{pred}(A_k)$ of its predecessors—called $A_k$'s *parents*—the event $A_k$ is conditionally independent of all other preceding events, or $\text{CI}[A_k, \mathbf{pa}(A_k), \mathbf{pred}(A_k) - \mathbf{pa}(A_k)]$. This structure can be depicted graphically as a *directed acyclic graph* (DAG): each event is a node in the graph, and there is a directed edge from node $A_j$ to node $A_k$ if and only if $A_j$ is a parent of $A_k$. We also refer to $A_k$ as the *child* of $A_j$. A DAG has no directed cycles and thus defines a partial order over its vertices. We assume without loss of generality that the event indices are consistent with this partial ordering; in other words, if $A_j$ is a predecessor of $A_k$ then $j < k$. We can write the joint probability distribution in a (usually) more compact form:

$$\Pr(A_1 A_2 \cdots A_M) = \prod_{k=1}^{M} \Pr(A_k | \mathbf{pa}(A_k)).$$

For each event $A_k$, we record a *conditional probability table* (CPT), which contains probabilities $\Pr(A_k | \mathbf{pa}(A_k))$ for all possible combinations of outcomes of events in $\mathbf{pa}(A_k)$. Thus, it is possible to implicitly represent the full joint with $O\left(M \cdot 2^{\max\{q(k)\}}\right)$ probabilities, instead of $2^M - 1$, where $q(k) = |\mathbf{pa}(A_k)|$ is the number of parents of $A_k$.

A *Markov independence* is a special type of conditional independence (Darroch et al., 1980; Pearl, 1988; Whittaker, 1990). The node $A_j$ and the set of nodes $X \subseteq Z - A_j$ are Markov independent, given another set $W \subseteq Z - X - A_j$, if $\text{CI}[A_j, W, X]$ and $A_j \cup W \cup X = Z$. Recall that $Z$ is the set of all modeled events.

A DAG is an *independency map*, or an *I-map*, of a probability distribution $\Pr$ if every independency implicit in the graph holds within $\Pr$ (Pearl, 1988). Note that a complete graph is a trivial I-map of any distribution over $\Omega$.

A DAG is *decomposable* if there is an edge between every two nodes that share a common child (Chyu, 1991; Darroch et al., 1980; Pearl, 1988; Shachter, Andersen, & Poh, 1991). Trees are a subset of decomposable DAGs, since every node has at most one parent. Complete graphs are also decomposable since *every* two nodes are connected. Any BN can be made decomposable by reorienting some edges and introducing new edges where needed (Chyu, 1991; Shachter et al., 1991). Though the decomposable representation can be exponentially larger than the original BN, it can still be exponentially more compact than the full joint distribution. The independencies encoded in a decomposable BN are all Markov independencies (Pearl, 1988).

## 3 EQUILIBRIUM IN A SECURITIES MARKET

### 3.1 EQUILIBRIUM AS CONSENSUS

The standard formulation of competitive equilibrium (3) is as a fixed point where each agent's demand is optimal at current prices, and each security's price balances aggregate demand. In this section, we examine an alternative characterization of equilibrium, recognized first by Drèze (1987). Agent $i$'s first-order condition for $x_i^{\langle j \rangle}$ is:

$$\frac{\partial U_i(\mathbf{x})}{\partial x_i^{\langle j \rangle}} = \sum_{\omega \in \Omega} \Pr_i(\omega) \frac{\partial u_i\left(\Upsilon_i^{\langle \omega \rangle}\right)}{\partial x_i^{\langle j \rangle}} = 0,$$

where $\Upsilon_i^{\langle \omega \rangle} = \sum_k \left(1_{\omega \in A_k} - p^{\langle k \rangle}\right) x_i^{\langle k \rangle}$ is its payoff in state $\omega$, and $1_{\omega \in A_k}$ is the indicator function that equals one if $\omega \in A_k$, and zero otherwise. Applying the chain rule

$$\sum_{\omega \in \Omega} \Pr_i(\omega) \left(1_{\omega \in A_j} - p^{\langle j \rangle}\right) u_i'\left(\Upsilon_i^{\langle \omega \rangle}\right) = 0$$

$$\sum_{\omega \in A_j} \Pr_i(\omega) u_i'\left(\Upsilon_i^{\langle \omega \rangle}\right)$$

$$- p^{\langle j \rangle} \sum_{\omega \in \Omega} \Pr_i(\omega) u_i'\left(\Upsilon_i^{\langle \omega \rangle}\right) = 0,$$

and solving for $p^{\langle j \rangle}$, we find that:

$$p^{\langle j \rangle} = \frac{\sum_{\omega \in A_j} \Pr_i(\omega) u_i'\left(\Upsilon_i^{\langle \omega \rangle}\right)}{\sum_{\omega \in \Omega} \Pr_i(\omega) u_i'\left(\Upsilon_i^{\langle \omega \rangle}\right)} = \Pr_i^{\text{RN}}(A_j). \quad (4)$$

In words, equilibrium can also be considered a fixed point where exchanges among agents induce a *consensus* on risk-neutral probabilities across available securities, and where the security prices themselves match these agreed-upon values.

### 3.2 COMPLETE MARKETS, COMPLETE CONSENSUS, AND PARETO OPTIMALITY

As described in Section 2.3, a securities market is *complete* when $S = |\Omega| - 1$ and all securities are *linearly independent*. In such a market, equilibrium allocations of risk are Pareto optimal: any gamble, contingent on *any* event $E \subseteq \Omega$, that is an acceptable purchase for one agent is *not* an acceptable sale for any other (Arrow, 1964).

A probability distribution over $\Omega$ has dimensionality $|\Omega| - 1$ (normalized likelihoods for the $|\Omega|$ states). Prices of securities in a complete market constitute $|\Omega| - 1$ linearly independent equations for these $|\Omega| - 1$ unknowns, and thus define unique probabilities for all states $\omega \in \Omega$, also called the *state prices* (Huang & Litzenberger, 1988; Varian, 1987). Denote these probabilities as $\Pr_0(\omega)$, and let $\Pr_0(E) = \sum_{\omega \in E} \Pr_0(\omega)$ be the price-probability of any event $E$, perhaps not directly corresponding to an available security.

The agents' risk-neutral distributions also have dimensionality $|\Omega| - 1$, subject to the $S$ constraints defined by (4). If the market is complete, it follows that $\Pr_i^{\text{RN}}$ is uniquely determined, and equals $\Pr_0$ for all $i$. That is, a complete market induces a compete consensus on risk-neutral probabilities. This suggests an intuitive explanation of why equilibrium allocations are Pareto optimal. All agents behave *as if* they are risk-neutral (payoff-maximizing) with identical beliefs. In such a situation, there are simply no differences of risk-preference or opinion on which to trade.

If $S < |\Omega| - 1$, then the consensus on risk-neutral probabilities is generally incomplete. Whenever $\Pr_h^{\text{RN}}(\omega) \neq \Pr_i^{\text{RN}}(\omega)$ for any $\omega$, there exists an acceptable exchange between agents $h$ and $i$, though perhaps not supported by the $S$ available securities. An equilibrium allocation in an incomplete market is not necessarily Pareto optimal.[1] But it *can* be, depending on the particular belief structures of the agents. Call a market *operationally complete* if its competitive equilibrium $(\mathbf{x}, \mathbf{p})$ is Pareto optimal (with respect to the agents involved), even if the market contains less than $|\Omega| - 1$ securities. As a degenerate example, an empty market is operationally complete for an economy of completely identical agents. Although such a market does not support all *conceivable* trades, it does support all *acceptable* trades among the given agents.

## 4 STRUCTURED MARKETS: AN ANALOGY TO BAYESIAN NETWORKS

Achieving completeness is, practically speaking, all but impossible; the required number of securities—exponential in the number of primary events—is simply too huge.

In attempting to represent probability distributions over $\Omega$, researchers in uncertain reasoning are faced with an analogous combinatorial explosion. The typical solution is to work with the factored event space, rather than the state

---

[1] Allocations are always efficient with respect to *available* securities, but not necessarily with respect to all states.

space, and to exploit any independencies among events using graphical models.

Continuing the analogy, securities markets can be structured according to the directed acyclic graph $D$ of any BN. Simply introduce one conditional security $\langle A_j | \mathbf{pa}(A_j)\rangle$ for every conditional probability $\Pr(A_j|\mathbf{pa}(A_j))$ in the network. For each event $A_j$ with $q(j) = |\mathbf{pa}(A_j)|$ parents, this adds $2^{q(j)}$ securities, one for each possible combination of outcomes of events in $\mathbf{pa}(A_j)$. Call such a market $D$-*structured*. Imagine for the moment that $D$ is fully connected (that is, no independencies are represented). Then a $D$-structured market contains $\sum_{j=1}^{M} 2^{j-1} = 2^M - 1 = |\Omega| - 1$ linearly independent securities, and is thus complete.

The benefit of a BN representation, and likewise a structured market, obtains when $D$ is less than fully connected, and thus the market contains less than $|\Omega| - 1$ securities. What can be said in this case? Certainly, depending on the beliefs and utilities of the agents, inefficient allocations are possible. Nonetheless, under circumstances explored below, the smaller market may suffice for operational completeness.

## 5 COMPACT MARKETS I

### 5.1 CONSENSUS ON RISK-NEUTRAL INDEPENDENCIES

Call a $D$-structured market a *risk-neutral independency market*, or an *RNI-market*, if, in equilibrium, $D$ is an I-map of $\Pr_i^{\text{RN}}$ for all agents $i$. That is, all agents' risk-neutral distributions agree with the independencies encoded in the market's structure. Paralleling our notation for true conditional independence, let $\text{CI}_i^{\text{RN}}[A_j, W, X]$ denote the risk-neutral conditional independence $\Pr_i^{\text{RN}}(A_j|WX) = \Pr_i^{\text{RN}}(A_j|W)$.

**Proposition 1** *At equilibrium in an RNI-market, $\Pr_h^{\text{RN}}(\omega) = \Pr_i^{\text{RN}}(\omega)$ for all agents $h, i$ and all states $\omega \in \Omega$.*

**Proof.** The market contains $\sum_{j=1}^{M} 2^{q(j)}$ securities, imposing an equal number of constraints on every agent's risk-neutral distribution via (4). For each event, I-mapness further imposes $2^{q(j)}(2^{j-1-q(j)} - 1)$ conditional independence constraints of the form $\text{CI}_i^{\text{RN}}[A_j, \mathbf{pa}(A_j), \mathbf{pred}(A_j) - \mathbf{pa}(A_j)]$, for all combinations of outcomes of events in $\mathbf{pa}(A_j)$ and all but one combination of outcomes of events in $\mathbf{pred}(A_j) - \mathbf{pa}(A_j)$ (the remaining one is implied by the others). Then every agent's risk-neutral distribution is subject to

$$\sum_{j=1}^{M} 2^{q(j)} + 2^{q(j)}(2^{j-1-q(j)} - 1)$$

$$= \sum_{j=1}^{M} 2^{j-1} = 2^M - 1 = |\Omega| - 1$$

identical, linearly independent constraints. Therefore $\Pr_h^{\text{RN}} = \Pr_i^{\text{RN}}$ for all $h, i$. $\square$

In an RNI-market, define the *state prices* $\Pr_0(\omega) = \Pr_i^{\text{RN}}(\omega)$ as the unique probabilities over $\Omega$ that are consistent with the prices of available securities and the independencies of $D$. The following corollary establishes that equilibrium prices for any of the $|\Omega| - 1 - S$ "missing" securities are also derivable from $\Pr_0$.

**Corollary 2** *Let $\langle p^{\langle 1 \rangle}, \ldots, p^{\langle S \rangle} \rangle$ be the equilibrium prices in an RNI-market. Introduce a new security $\langle E \rangle$. Then $\langle p^{\langle 1 \rangle}, \ldots, p^{\langle S \rangle}, \Pr_0(E)\rangle$ are equilibrium prices in the expanded market.*

**Proof.** Before the extra security is introduced, all agents' risk-neutral probabilities $\Pr_i^{\text{RN}}(E)$ already equal $\Pr_0(E)$, without buying or selling any quantity of the security. It follows that, with the additional security, the equilibrium condition (4) is satisfied with $x_i^{\langle E \rangle} = 0$ for all $i$, $p^{\langle E \rangle} = \Pr_0(E)$, and all other prices unchanged. $\square$

The number of securities in an RNI-market, $O\left(M \cdot 2^{\max\{q(j)\}}\right)$, can be exponentially smaller than the $2^M - 1$ required for traditional completeness. The following corollary shows that the more compact market supports allocations that are equally efficient.

**Corollary 3** *Every RNI-market is operationally complete. That is, the equilibrium allocations $\mathbf{x}$ and state prices $\Pr_0$ in an RNI-market constitute an equilibrium in a (truly) complete market composed of the same agents.*

**Proof.** By repeated application of Corollary 2, we can add the $|\Omega| - 1 - S$ securities necessary to complete the market.[2] For each new security, a price consistent with $\Pr_0$, coupled with zero demand from all agents, satisfies (4). All complete markets, regardless of structure, support the same equilibrium allocations and state prices (Huang & Litzenberger, 1988; Mas-Colell et al., 1995; Varian, 1987). $\square$

Proposition 1 and its corollaries are equilibrium results only. We sketch here one possible procedure for *reaching agreement on the market structure*.[3] Begin with securities in only the $M$ events: $\langle A_1 \rangle, \ldots, \langle A_M \rangle$. If any agent's demand for $\langle A_k | A_j \rangle$ (for any $j < k$) at price $p^{\langle A_k \rangle}$ is nonzero, then it creates a new market in $\langle A_k | A_j \rangle$. If, at some future time, the agent has zero demand for its new security, then it may retract the security. An additional condition for equilibrium is that no agent desires to create or withdraw any markets. Then, in equilibrium, it should be the case that all agents' risk-neutral independencies agree with the market structure, and that the market is operationally complete. We might want to add a transaction cost for opening new markets, so that equilibrium only ensures that risks are hedged up to a threshold cost.

---

[2] A natural set to add are the $\sum_{j=1}^{M} 2^{q(j)}(2^{j-1-q(j)} - 1)$ securities of the form $\langle A_j | \mathbf{pred}(A_j)\rangle$, for all events $A_j$, all combinations of outcomes of $\mathbf{pa}(A_j)$, and all but one combination of outcomes of $\mathbf{pred}(A_j) - \mathbf{pa}(A_j)$.

[3] This procedure is similar to Geiger's (1990) protocol for eliciting independence structures from experts.

## 5.2 COMPUTATIONAL COMPLEXITY OF ARBITRAGE

Imagine that, after equilibrium is reached in an RNI-market, a redundant security is introduced, say $\langle A_M \rangle$. The equilibrium price of $\langle A_M \rangle$ is already determined (Corollary 2): it must equal $\text{Pr}_0(A_M) = \text{Pr}_i^{\text{RN}}(A_M)$. Furthermore, if the current price does *not* equal $\text{Pr}_0(A_M)$, then the market is not in equilibrium, and arbitrage is possible. For example, if $p^{\langle A_M \rangle} < \text{Pr}_0(A_M)$, then an outside observer $O$ could purchase it at the going price and sell it to any of the agents at price $p^*$ such that $p^{\langle A_M \rangle} < p^* < \text{Pr}_i^{\text{RN}}(A_M) = \text{Pr}_0(A_M)$. Although $O$ does not have direct access to $\text{Pr}_0(A_M)$, it is uniquely computable given the other prices and the independence structure of $D$.

If $O$ can find an arbitrage opportunity by correctly pricing the redundant security, then $O$ can perform Bayesian inference, which is #P-complete (Cooper, 1990).

## 6 COMPACT MARKETS II: CONSENSUS ON TRUE INDEPENDENCIES

Equilibrium agreement on risk-neutral independencies may seem a somewhat strange condition, especially considering that the $\text{Pr}_i^{\text{RN}}$ are changing as transactions occur. Some authors argue that, since agents appear to act according to $\text{Pr}_i^{\text{RN}}$ and standard elicitation techniques reveal $\text{Pr}_i^{\text{RN}}$, risk-neutral probabilities are in fact no less "real" than true probabilities (Kadane & Winkler, 1988; Nau & McCardle, 1991). However, while it seems reasonable that agents would have true independencies in common (Pearl, 1993; Smith, 1990), it is harder to justify why their risk-neutral independencies would coincide. This section develops a theory of compact markets based on consensus on true independencies. If, despite any quantitative differences between $\text{Pr}_i$ and $\text{Pr}_i^{\text{RN}}$, an agent's true independencies were always manifest as risk-neutral independencies, then results concerning RNI-markets would carry over unchanged. Section 6.1 demonstrates that this is indeed the case for a subclass of agents and a subset of independencies. Section 6.2 discusses how known limitations of belief aggregation procedures restrict the possibility of obtaining compact markets under more general circumstances.

### 6.1 CONSENSUS ON MARKOV INDEPENDENCIES

A commonly assumed risk-averse utility form is *exponential utility*: $u_i(\mu) = -e^{-c_i\mu}$. This utility form is synonymous with *constant absolute risk aversion* (CARA), where $c_i$ is agent $i$'s coefficient of risk aversion, or $1/c_i$ its risk tolerance. As the agent's wealth increases, its marginal utility for unit dollars decreases (since it is risk-averse), but the *amount* of its aversion to risk remains constant at $c_i$.

In this section, we show that, in economies composed of agents with CARA, markets structured according to agreed upon (true) *Markov* independencies are operationally complete. Define an *independency market*, or an *I-market*, as a $D$-structured market such that $D$ is an I-map of $\text{Pr}_i$ for all agents $i$ (i.e., all agents' true distributions agree with the independencies in $D$). An I-market is *decomposable* if $D$ is decomposable—every node's parents are fully connected.

Let $Z = \{A_1, \ldots, A_M\}$ be the set of all events, $A_j \in Z$ a particular event, and $W \subseteq Z - A_j$ and $X = Z - W - A_j$ subsets of events. We are interested in whether agent $i$'s Markov independencies $\text{CI}_i[A_j, W, X]$ are reflected as a risk-neutral independencies $\text{CI}_i^{\text{RN}}[A_j, W, X]$, and are thus observable. For brevity, we drop the subscript $i$ when only one agent is under consideration.

**Proposition 4**

$$\text{CI}[A_j, W, X] \& \left( \frac{u'\left(\Upsilon^{(A_j WX)}\right)}{u'\left(\Upsilon^{(\bar{A}_j WX)}\right)} = \frac{u'\left(\Upsilon^{(A_j W\tilde{X})}\right)}{u'\left(\Upsilon^{(\bar{A}_j W\tilde{X})}\right)} \right)$$
$$\Rightarrow \text{CI}^{\text{RN}}[A_j, W, X], \qquad (5)$$

*where the second precondition must hold for all possible joint outcomes of the events in $W$, and all pairs $(X, \tilde{X})$ of different joint outcomes of events in $X$.*

**Proof.**

$$\frac{u'\left(\Upsilon^{(A_j WX)}\right)}{u'\left(\Upsilon^{(\bar{A}_j WX)}\right)} = \frac{u'\left(\Upsilon^{(A_j W\tilde{X})}\right)}{u'\left(\Upsilon^{(\bar{A}_j W\tilde{X})}\right)}$$

$$\text{Pr}(A_jW) + \text{Pr}(\bar{A}_jW)\frac{u'\left(\Upsilon^{(A_j WX)}\right)}{u'\left(\Upsilon^{(\bar{A}_j WX)}\right)}$$
$$= \text{Pr}(A_jW) + \text{Pr}(\bar{A}_jW)\frac{u'\left(\Upsilon^{(A_j W\tilde{X})}\right)}{u'\left(\Upsilon^{(\bar{A}_j W\tilde{X})}\right)}$$

$$= \frac{\frac{\text{Pr}(A_jW)\text{Pr}(WX)}{\text{Pr}(W)}u'\left(\Upsilon^{(A_j WX)}\right)}{\frac{\text{Pr}(A_jW)\text{Pr}(WX)}{\text{Pr}(W)}u'\left(\Upsilon^{(A_j WX)}\right) + \frac{\text{Pr}(\bar{A}_jW)\text{Pr}(WX)}{\text{Pr}(W)}u'\left(\Upsilon^{(\bar{A}_j WX)}\right)}$$
$$= \frac{\frac{\text{Pr}(A_jW)\text{Pr}(W\tilde{X})}{\text{Pr}(W)}u'\left(\Upsilon^{(A_j W\tilde{X})}\right)}{\frac{\text{Pr}(A_jW)\text{Pr}(W\tilde{X})}{\text{Pr}(W)}u'\left(\Upsilon^{(A_j W\tilde{X})}\right) + \frac{\text{Pr}(\bar{A}_jW)\text{Pr}(W\tilde{X})}{\text{Pr}(W)}u'\left(\Upsilon^{(\bar{A}_j W\tilde{X})}\right)}$$

$$\frac{\text{Pr}(A_jWX)u'\left(\Upsilon^{(A_j WX)}\right)}{\text{Pr}(A_jWX)u'\left(\Upsilon^{(A_j WX)}\right) + \text{Pr}(\bar{A}_jWX)u'\left(\Upsilon^{(\bar{A}_j WX)}\right)}$$
$$= \frac{\text{Pr}(A_jW\tilde{X})u'\left(\Upsilon^{(A_j W\tilde{X})}\right)}{\text{Pr}(A_jW\tilde{X})u'\left(\Upsilon^{(A_j W\tilde{X})}\right) + \text{Pr}(\bar{A}_jW\tilde{X})u'\left(\Upsilon^{(\bar{A}_j W\tilde{X})}\right)}$$

$$\frac{\text{Pr}^{\text{RN}}(A_jWX)}{\text{Pr}^{\text{RN}}(A_jWX) + \text{Pr}^{\text{RN}}(\bar{A}_jWX)} = \frac{\text{Pr}^{\text{RN}}(A_jW\tilde{X})}{\text{Pr}^{\text{RN}}(A_jW\tilde{X}) + \text{Pr}^{\text{RN}}(\bar{A}_jW\tilde{X})}$$

$$\text{Pr}^{\text{RN}}(A_j|WX) = \text{Pr}^{\text{RN}}(A_j|W\tilde{X})$$

□

The second precondition in (5) says that the ratio of marginal utility in states where $A_j$ does not occur to marginal utility in states where $A_j$ does occur cannot depend of the outcomes of events in $X$. This is true (and indeed $\Pr^{RN} = \Pr$) if the agent's marginal utility $u'$ is constant across states. This holds if the agent is risk neutral, and holds approximately if utility is state-independent and $\Upsilon^{\langle \omega_j \rangle} \approx \Upsilon^{\langle \omega_k \rangle}$. But this approximation is not realistic for an agent engaged in trading securities, since a central role of the market is precisely to enable the transfer of wealth across states.

Let $\Upsilon^{\langle A_j W \rangle}$ be the agent's payoff from all securities that depend only the outcomes of events in $A_j \cup W$. Examples are $\langle A_j \rangle$, $\langle A_j W \rangle$, and $\langle A_j | W \rangle$, which return the same dollar amount regardless of the realizations of events in $X = Z - W - A_j$. Similarly, let $\Upsilon^{\langle WX \rangle}$ be the payoff from securities that do not depend on $A_j$.

Suppose that the agent exhibits CARA, and that its payoffs are separable according to $\Upsilon^{\langle A_j WX \rangle} = \Upsilon^{\langle A_j W \rangle} + \Upsilon^{\langle WX \rangle} - \Upsilon^{\langle W \rangle}$. Separability essentially means that any of the agent's securities (or prior stakes) whose payoff depends on $A_j$ cannot also depend on events in $X$. In this case,

$$\frac{u'\left(\Upsilon^{\langle A_j WX \rangle}\right)}{u'\left(\Upsilon^{\langle \bar{A}_j WX \rangle}\right)} = \frac{u'\left(\Upsilon^{\langle A_j W \rangle} + \Upsilon^{\langle WX \rangle} - \Upsilon^{\langle W \rangle}\right)}{u'\left(\Upsilon^{\langle \bar{A}_j W \rangle} + \Upsilon^{\langle WX \rangle} - \Upsilon^{\langle W \rangle}\right)}$$
$$= \frac{ce^{-c\Upsilon^{\langle A_j W \rangle}}e^{-c\Upsilon^{\langle WX \rangle}}e^{c\Upsilon^{\langle W \rangle}}}{ce^{-c\Upsilon^{\langle \bar{A}_j W \rangle}}e^{-c\Upsilon^{\langle WX \rangle}}e^{c\Upsilon^{\langle W \rangle}}}$$
$$= \frac{ce^{-c\Upsilon^{\langle A_j W \rangle}}e^{-c\Upsilon^{\langle W\tilde{x} \rangle}}e^{c\Upsilon^{\langle W \rangle}}}{ce^{-c\Upsilon^{\langle \bar{A}_j W \rangle}}e^{-c\Upsilon^{\langle W\tilde{x} \rangle}}e^{c\Upsilon^{\langle W \rangle}}}$$
$$= \frac{u'\left(\Upsilon^{\langle A_j W \rangle} + \Upsilon^{\langle W\tilde{x} \rangle} - \Upsilon^{\langle W \rangle}\right)}{u'\left(\Upsilon^{\langle \bar{A}_j W \rangle} + \Upsilon^{\langle W\tilde{x} \rangle} - \Upsilon^{\langle W \rangle}\right)} = \frac{u'\left(\Upsilon^{\langle A_j W\tilde{x} \rangle}\right)}{u'\left(\Upsilon^{\langle \bar{A}_j W\tilde{x} \rangle}\right)}$$

Thus the constraint on utility in (5) is satisfied, and any Markov independencies are observable.

We are now in a position to derive the main result of this section.

**Proposition 5** *When all agents have CARA, every decomposable I-market is an RNI-market.*

**Proof.** Let $W_j$ be the set of direct parents and direct children of event $A_j$, and $X_j$ all other events. From decomposability and I-mapness, we can infer that

1. $CI_i[A_j, W_j, X_j]$ for all agents $i$ and events $j$,

2. none of the securities $\langle A_j | \mathbf{pa}(A_j) \rangle$ that are contingent on $A_j$ depend on $X_j$, and

3. none of the securities $\langle A_k | \mathbf{pa}(A_k) \rangle$ such that $A_j \in \mathbf{pa}(A_k)$ that are conditional on $A_j$ depend on $X_j$.

Items 2 and 3 ensure separability of payoffs from the available securities (we assume that any prior stakes are also separable). Then, invoking Proposition 4, $CI_i^{RN}[A_j, W_j, X_j]$ for all agents $i$ and events $j$. As a result, $D$ is an I-map of every $\Pr_i^{RN}$ regardless of allocations or prices, including those at equilibrium. □

Proposition 1 and Corollaries 2 and 3 are immediately applicable. In particular, for agents with CARA, every decomposable I-market is operationally complete.

### 6.2 INHERENT LIMITATIONS

One might wonder whether compact I-markets are possible for larger classes of agents or independencies. It can be shown via counterexample that, even when all agents have CARA, a market conforming to agreed-upon (possibly non-Markov) independencies will not always be operationally complete. Moreover, when all agents have logarithmic utility for money (another commonly assumed utility form), even a market conforming to agreed-upon Markov independencies will not always be operationally complete.

Although we do not have a formal statement of impossibility, results from statistical belief aggregation suggest that agreement on true independencies will *not* be sufficient in general to yield compact and operationally complete markets. The state prices $\Pr_0$ in a securities market are a function of all the agents' beliefs (and their utilities), and as such essentially constitute a measure of aggregate belief. Many researchers have studied belief aggregation functions (Genest & Zidek, 1986), and several impossibility theorems severely restrict the class of functions that preserve unanimously held independencies (Genest & Wagner, 1987), even when restricted to independencies among the primary events (Pennock & Wellman, 1999). The aggregation "function" of a securities market is of course subject to the same limitations. We suspect that, for many configurations of agents, markets structured according to unanimously-held true independencies will not yield provably optimal allocations of risk. Nevertheless, it may well be the case that structured markets can yield *approximately* optimal allocations over a wider range of agent populations.

## 7 CONCLUSIONS

Rational risk-averse agents will seek ways to mitigate the dangers inherent in an uncertain world by reducing their exposure to risk. Whenever two agents exhibit divergent tolerances for risk (e.g., an insurance company and a homeowner), or disagree on the likelihood of world outcomes (e.g., a bettor on St. Louis to win Super Bowl XXXIV and a bettor on Tennessee), there may be an opportunity for an exchange of state-contingent wealth—essentially a portfolio of *securities*—that both agents deem beneficial. To guarantee that all desirable exchanges of risk are supported, a market must be *complete*, or contain at least $2^M - 1$ linearly independent securities, where $M$ is the number of relevant uncertain events. Clearly, this number of securities is prohibitive in even modestly complex domains.

In this paper, we showed that the same principles used to succinctly represent joint probability distributions can aid in reducing the required number of securities. We illustrated how markets can be structured analogously to Bayesian networks. We derived two conditions under which compact markets—in some cases with exponentially fewer securities than complete markets—can still support all desirable exchanges of risk. The most general condition is that all agents' risk-neutral independencies agree with the independencies encoded in the market's structure. For populations of agents with constant absolute risk aversion, agreement on Markov independencies is sufficient.

We plan to evaluate empirically whether structured markets can yield efficiency gains even when agents do not meet all of these theoretical sufficiency requirements. As a potential future application, one might imagine structuring a set of derivatives so as to increase opportunities for agents to hedge their risks, while at the same time keeping the number of financial instruments required at a minimum.

### Acknowledgments

Thanks to Eric Horvitz, C. Lee Giles, the members of the Decision Machine Research Group at the University of Michigan, and the anonymous reviewers. This work was partially supported by AFOSR Grant F49620-97-0175.